\documentclass[11pt]{article}
\usepackage{epsfig,amscd,amssymb}
\usepackage{graphicx}
\usepackage{latexsym}
\usepackage{amssymb,amsfonts}
\usepackage{amsmath} 
\begin{document}

%
%
%
%
\title{Cooper pairs and exclusion statistics \\ 
from coupled free-fermion chains}%

\author{Paul Fendley$^1$ and Kareljan Schoutens$^2$ 
\medskip \\ 
$^1$ Department of Physics, University of Virginia, \\
Charlottesville, VA 22904-4714 USA  
\medskip \\  
$^2$ Institute for Theoretical Physics,  
University of Amsterdam,\\
 Valckenierstraat 65, 1018 XE Amsterdam, The Netherlands \\  
}
\smallskip 

\date{January 29, 2007}
\maketitle

\begin{abstract}
We show how to couple two free-fermion chains so that the excitations
consist of Cooper pairs with zero energy, and free particles obeying
(mutual) exclusion statistics. This behavior is reminiscent of anyonic
superconductivity, and of a ferromagnetic version of the
Haldane-Shastry spin chain, although here the interactions are
local. We solve this model using the nested Bethe ansatz, and find all
the eigenstates; the Cooper pairs correspond to exact-string or
``0/0'' solutions of the Bethe equations. We show how the model
possesses an infinite-dimensional symmetry algebra, which is a
supersymmetric version of the Yangian symmetry algebra for the
Haldane-Shastry model.
\end{abstract}

\section{Introduction}

The Bethe ansatz is one of the few reliable methods physicists have
for probing properties of systems with strong correlations. For the
systems to which it is applicable, it gives a
way of doing exact computations. 

One of the most valuable insights it gives is in the quasiparticle
spectrum of one-dimensional quantum systems. A number of profound but
highly non-obvious properties have been discovered using the Bethe
ansatz. For example, Bethe's original paper gives the tools necessary
to show that the quasiparticles in the one-dimensional
antiferromagnetic Heisenberg model are gapless and have a linear
dispersion relation \cite{Bethe31}.  It took more than half a century
and some novel non-perturbative physics to understand that these
excitations have spin $1/2$, and to explain this in terms of field
theory \cite{Faddeev,Haldane85}. After the Bethe ansatz was generalized to
systems with multiple types of particles via the invention of the
nested Bethe ansatz \cite{Yang67}, it was subsequently shown how the
spectrum of the one-dimensional Hubbard model is spin-charge
separated. Even though the Hubbard model is comprised of electrons
with both charge and spin, the quasiparticles in one dimension have
either spin or charge, but not both \cite{Lieb68}. It was again quite
some time before this result was understood in depth from the
field-theory point-of-view.

In this paper we use the Bethe ansatz to find and describe a 
strongly-correlated one-dimensional system with two kinds of
interesting excitations. The system consists of two coupled
free-fermion chains; the excitations are Cooper pairs and
exclusons. 

Cooper pairs of course are familiar from studies of superconductivity
and fermionic superfluidity. They are comprised of two fermions bound
(in momentum space) into a state which has zero energy. The Cooper
pairs here arise in a fashion very reminiscent of anyonic
superconductivity \cite{Laughlin88}. Anyons are quasiparticles in two
dimensions which have statistics generalizing those of fermions and
bosons: they pick up a phase under exchange, instead of a sign. One
kind of interaction between our fermions is a one-dimensional analog:
a fermion on one chain picks up a phase $\pm i$ when it moves past a
fermion on the other chain.

Exclusons are also a generalization of the idea of fermions and bosons
\cite{Haldane91b}. However, as opposed to anyons, they can occur in
one (or any) dimension. The usual notion of statistics is not
applicable in one dimension because particles can not be exchanged
without coming close to each other. Nevertheless, in any dimension
fermions can be defined as particles which are allowed at most one to
a level, and bosons those where any number is allowed.  Exclusion
statistics describe more general rules for how particles fill energy
levels. In our model, we will show that there are two types of
exclusons, associated with each of the two chains. Each of the two
types has their own (identical) set of energy levels, which with
appropriate boundary conditions are those of free particles. Amongst
themselves, each type behaves like a fermion, allowing at most one
quasiparticle per level. However, when say an excluson of type $1$ is
in a given energy level, not only does it forbid another of type $1$
from having that energy, but it also forbids a particle of type $2$
from having that energy as well.

Exclusons in one dimension are well known to occur as quasiparticle
excitations of the Haldane-Shastry chain, which is a variant of the
Heisenberg model with long-range interactions \cite{Haldane88}.  In
both our model and the Haldane-Shastry chain \cite{Haldane91a}, a nice
property is that the interaction between exclusons is purely
statistical. In other words, the energy levels are those of free
particles: the only interactions are in the rules for filling these
levels. It is important to note that it is the quasiparticles which
have only exclusion interactions: the original degrees of freedom in
both cases are strongly interacting. One very interesting distinction
between our model and the Haldane-Shastry chain is that the
interactions here are nearest-neighbor.

Even after 75 years, computations using the Bethe ansatz are usually
quite technical, and this one is no exception. However, we find it
remarkable that here the technical complications result in elegant
physics. For example, to obtain our results we impose open boundary
conditions, which require much more effort to deal with in the Bethe
ansatz than periodic ones do. Nevertheless, not only does
the Bethe ansatz still work here, but the fact eigenstates are
standing waves ends up making it possible to find them all in closed
form. This is how we manage to show that this strongly-interacting
system has quasiparticle excitations which are free exclusons and
Cooper pairs.

The Cooper pairs themselves arise as a consequence of resolving
another technical complication: they turn out to be
``exact-string'' solutions of the Bethe equations. The Bethe equations
are coupled polynomial equations whose solutions determine the allowed
momenta in the wavefunction. String solutions are those for which some
of these polynomials vanish in the thermodynamic limit; they occur
even in the Heisenberg model discussed in Bethe's original paper. In
an exact-string solution, they vanish even for a finite number of
sites. In the way the Bethe equations are usually written, this is a
singular ``0/0'' limit: both the numerator and the denominator of a
ratio are exactly zero. However, it is explained in detail in
\cite{Baxter} how to define the
Bethe equations so that exact-string solutions are not singular. Including
such solutions is necessary to ensure that all eigenstates of the
Hamiltonian follow from the Bethe ansatz.

Moreover, the presence of exact-string solutions in the XXZ spin chain
described in \cite{Baxter} is closely related to the appearance of a
very interesting extended-symmetry algebra there \cite{McCoy}. We find
an analogous algebra for our model as well. Because our model turns
out to have a supersymmetry, this extended symmetry algebra is an
extension of the supersymmetry algebra. A similar symmetry algebra
called the Yangian appears in the Haldane-Shastry chain
\cite{HSYangian}.

In section \ref{sec:results}, we present our model and describe in
detail the results of this paper. In section \ref{sec:bethe}, we
derive these results using the Bethe ansatz. In section
\ref{sec:susy}, we discuss the extended symmetry algebra, and its
connection to supersymmetry. We present our conclusions in section \ref{sec:conclusion}.

\section{The model and the results}
\label{sec:results}

We consider two free-fermion chains, each with $L$ sites. It is
convenient to label the sites on one chain by odd integers
$1,3,\dots,2L-1$ and on the other by even integers $2,4,\dots,2L$.
The spinless fermions on 
the two chains are created by the operators $c_j^\dagger$ 
which anticommute except for $\{c_k,c_j^{\dagger}\}=\delta_{jk}$. 
The fermions on
a given chain do not interact among themselves, except of course for the
requirement that two fermions cannot occupy the same site.
The number operator $n_j=c^\dagger_j c_j$ is 1 if the site $j$ is
occupied and $0$ if it is unoccupied. 
We allow hopping along each
chain, but not from one chain to the other, so the fermion-number operators
$$F_1= \sum_{k=1}^L n_{2k-1}\ , \qquad\quad
F_2= \sum_{k=1}^L n_{2k}
$$
are each conserved.
A state with fermion numbers $f_1$ and $f_2$ on the
chains is of the form
\begin{equation}
\sum_{j_1\dots j_{f_1+f_2}} \varphi\left(j_1,\dots,j_{f_1}|j_{f_1+1},\dots
j_{f_1+f_2}\right)
c^\dagger_{j_1}\dots c^\dagger_{j_{f1+f2}}|0\rangle\ ,
\label{estates}
\end{equation}
where $|0\rangle$ is the empty state. Here and for the remainder of
the paper, $j_a$ is odd if $a\le f_1$ and even if $a> f_1$.

Let us first consider two decoupled free-fermion chains with open
boundary conditions.
We include hopping, a specific chemical potential, and a boundary magnetic
field, so that the Hamiltonian is 
\begin{equation}
H_f
= 2F_1 + 2F_2  +
\sum_{j=2}^{2L-1}
\left[ 
c_{j+1}^\dagger  c_{j-1} + 
c_{j-1}^\dagger  c_{j+1}\right]
+ H_{b}\ .
\label{Ham}
\end{equation}
The specific boundary magnetic field we choose gives
\begin{equation}
H_b= -n_1-n_{2L}\ .
\label{bcmag}
\end{equation}
so that it lowers the energy when a particle is on one end of each chain.

An elementary computation gives the eigenstates of
$H_f$ to be of the
form (\ref{estates}) with
\begin{equation}
\varphi_{\hbox{free}}
= \prod_{a=1}^{f_1+f_2} \sin(p_a j_a)
\label{phifree}
\end{equation}
where
\begin{equation}
p_a = m_a \frac{\pi}{2L+1}
\label{pfree}
\end{equation}
with $m_a$ an integer from $1\dots L$. Even though the open boundary
conditions mean that the system is not translation-invariant and there
is no overall momentum conservation, we still refer to the $p_a$ as
the momenta.  The shift in the denominator of (\ref{pfree}) is due to
the boundary magnetic field.  The energy of this eigenstate is
\begin{equation}
E= \sum_{a=1}^{f_1+f_2} (2+2\cos(2p_a)). 
\label{Ep}
\end{equation}
To impose Fermi statistics and forbid double occupancy, one sums over
these eigenstates as in (\ref{phif0}) below. One can have $m_a=m_b$
only if $a=b$ or if $a$ and $b$ label particles on different chains,
i.e.\ $a\le f_1$ and $b>f_1$. For a given $f_1$ and $f_2$ there are
therefore $\binom{L}{f_1}\binom{L}{f_2}$ different choices of
momentum. This is the number of states, so (\ref{phifree}) is a complete
set of states.

We now couple the two chains in a zig-zag fashion, so that each site
$j$ is adjacent to the sites $j\pm 1$ on the other chain.  There are
two kinds of interaction between the chains. The first is simply an
attractive interaction between nearest-neighbor fermions on different
chains. The second is a phase factor $\pm i$ picked up when a fermion
on one chain hops past one on the other, which can be thought of as
resulting from flux tubes attached to the particles. The Hamiltonian
is then $H_f + H_c$, where
\begin{equation}
H_c= 
-\sum_{j=2}^{2L-1} 
\left[(1-i)c_{j+1}^\dagger n_j  c_{j-1} + 
(1+i)c_{j-1}^\dagger n_j c_{j+1} \right] 
\ -\ 2\sum_{j=1}^{2L-1} n_jn_{j+1}\ .
\label{Hc}
\end{equation}
The specific values of the couplings in (\ref{Ham}) and (\ref{Hc}) are
chosen because they result in a model with a great deal of symmetry,
which we will discuss in depth in this paper. Some of the symmetries
(e.g.\ integrability) persist under certain deformations of the
couplings, but we will confine our studies to this specific
Hamiltonian.

The model with Hamiltonian $H_f+H_c$ is solvable using the nested
Bethe ansatz, and in the next section we present the solution in
detail. The Bethe equations can be solved explicitly, and the
resulting physics is quite elegant. We therefore discuss the key
physical results here. 

We find that in this strongly-coupled theory, the energy levels are
still those of free fermions, i.e.\ they are given by (\ref{Ep}) with
momenta quantized in the free-fermion form (\ref{pfree}). However, the
theory is {\em not} the same as that of the decoupled chains: the
strong interactions have several interesting effects. The
wavefunctions can still be written in a form quite similar to
(\ref{phifree}), so that each state can be characterized by a set of
momenta $p_a$ with $a=1,\dots, f_1+f_2$. The energy is still given by
(\ref{Ep}). However, not all momenta obey (\ref{pfree}) or are even
real. A pair of particles, one from each chain, can form a {\em Cooper
pair}. Namely, two of the momenta (say $p_1$ and $p_2$) obey
\begin{equation}
\cos^2(p_1) = -\cos^2(p_2).
\label{pcp}
\end{equation}
This existence of Cooper pairs does not contradict the first sentence
of this paragraph, because each pair obeying (\ref{pcp}) has no net
contribution to the energy. If $f_1=f_2$, then one can pair all the
particles and obtain a state with energy zero. We show on general
grounds in section \ref{sec:susy} that $E\ge 0$, so a state comprised
entirely of Cooper pairs is a ground state.

Any momenta not part of a Cooper pair must obey (\ref{pfree}), so the
energy levels are indeed those of free fermions. The degeneracy of
each level, however, is not the same as that for the decoupled
chains. One can add Cooper pairs without changing the energy, which
obviously will change the degeneracies. Even in a state with no Cooper
pairs, there is a crucial difference.  In both coupled and decoupled
chains, the numbers $f_1$ and $f_2$ of fermions on each chain are
conserved. When the chains are decoupled, two momenta can be the same,
if they are associated with particles on different chains. When the
chains are coupled via $H_c$, we find that {\em no two momenta can be equal.}
In fact, the individual momenta cannot be associated with one chain or
the other like they can in the decoupled case: for $C$ Cooper pairs,
the wavefunction is labeled by $f_1+f_2-2C$ distinct momenta obeying
(\ref{pfree}), and $C$ pairs of momenta obeying (\ref{pcp}).

Thus we can interpret the states with non-zero energy
in terms of quasiparticles with
momentum (\ref{pfree}). We call these quasiparticles exclusons for the
following reason.
Consider fixed numbers $f_1$ and $f_2$ of
fermions on each chain, and also a fixed number of Cooper pairs
$C$. There are two types of exclusons: $N_1=f_1-C$ of type 1, and
$N_2=f_2-C$ of type 2. The energy of this state is given by
\begin{equation}
E=\sum_{a=1}^{N_1+N_2} 4\cos^2(m_a\pi/(2L+1))\ 
\label{Eexc}
\end{equation}
with the $m_a \in \{1,\dots L\}$ distinct integers.
We show in section \ref{sec:bethe} that the number of
states with this energy is
\begin{equation}
d_E= \binom{N_1+N_2}{N_1} \binom{L-N_1-N_2}{C}
\label{degen}
\end{equation}
We have checked this formula for the degeneracy
directly by numerically diagonalizing the Hamiltonian (which of course
is how we originally discovered it).

In the quasiparticle interpretation of (\ref{Eexc},\ref{degen}), the first
binomial arises from the exclusons, the second from the Cooper
pairs. The $N_i$ exclusons of each type are indistinguishable from
each other, so the number of ways of assigning $N_1+N_2$ momenta to
them which result in distinct wavefunctions is
$\binom{N_1+N_2}{N_1}$. The interpretation of the second piece is that
the Cooper pairs can be thought of indistinguishable quasiparticles like
the exclusons, and there is one possible Cooper pair for each allowed
momentum.  Moreover, the presence of an excluson with a given momentum
prohibits a Cooper pair from occupying the corresponding level.

The exclusons, the quasiparticles of non-zero energy, obey mutual {\em
exclusion statistics} \cite{Haldane91b}. Even though in our counting we
have treated exclusons on chain 1 and on chain 2 as distinct species,
only one can fill a given momentum level. This is thus a generalized
exclusion principle: the presence of one kind of quasiparticle affects
how others fill levels. A mutual exclusion principle is typical for
systems solvable by the nested Bethe ansatz; what is novel here is
that except for this mutual exclusion, the quasiparticles are
otherwise blind to each other's presence. Namely, the energy levels
are unaffected by how many quasiparticles are present: the only
effects are on how the levels fill.

This mutual exclusion principle involves the Cooper pairs as
well. Fermions of the same type must still of course obey Pauli
exclusion. Thus the presence of exclusons restricts the allowed
momenta of the Cooper pairs, and reduces the size of the space of
states for the Cooper pairs as well.

One important issue to note is that the momentum of one of the
particles in a Cooper pair is arbitrary, as long as it is not the same
as any of the exclusons. The arbitrariness is due to the degeneracy:
any linear combination of the $d_E$ states with fixed excluson momenta
is an eigenstate of the Hamiltonian. This arbitrariness in choosing
certain momenta in the Bethe ansatz is discussed in depth in
\cite{Baxter}, and we will return to this issue in section
\ref{sec:bethe}. The counting rules become clearer if we choose any
arbitrary momenta to be one of the values obeying (\ref{pfree}) which
are not already present. The momentum of the second particle in each
Cooper pair is then fixed by (\ref{pcp}). The second factor in the
degeneracy (\ref{degen}) then simply corresponds to the number of
different such choices.

The factor $d_E$ looks somewhat complicated. However, it ends up
yielding an extremely simple and elegant formula for the {\em grand}
canonical partition function. The allowed momentum values
(\ref{pfree}) are independent of how many quasiparticles are present,
so we can treat the grand canonical partition function as a product
over momentum levels in the usual free-particle fashion. The reason
this is possible here is that the way to fix the arbitrariness in the
Cooper-pair momenta uniquely is to make it obey the same rules as the
exclusons do. In other words, there are four possibilities for each of
these momentum levels: it is empty, occupied by an excluson on chain 1
or by an excluson on chain 2, or occupied by a Cooper pair. The first
and last of these possibilities have energy $0$, whereas an excluson
in the $a$th momentum level has energy $4\cos^2(p_a)$. If $\lambda_1$
and $\lambda_2$ are the fugacities for each fermion on chains $1$
and $2$ (not including the chemical potentials of $-2$ already in
(\ref{Ham})), the grand canonical partition function at temperature
$T$ is then
\begin{equation}
Z = \prod_{m=1}^L 
\left[1+\lambda_C + (\lambda_1+\lambda_2)e^{-4\cos^2(m\pi/(2L+1))/T}\right]
\label{grandcp}
\end{equation}
where $\lambda_C=\lambda_1\lambda_2$. In the limit $T\to\infty$, $Z$
simply counts the number of allowed states. This limit indeed yields
the correct $Z\to 4^L=2^{2L}$ for two chains of length $L$ (each site is
either occupied or unoccupied). From (\ref{grandcp}) one can compute
any thermodynamic quantity, such as the specific heat, or the density
of Cooper pairs.

One can easily recover the degeneracies in (\ref{degen}) by
expanding out the product in $Z$, i.e.
$$Z = \sum_{{\cal N}=0}^L\sum_{N_1=0}^{\cal N} \sum_{C=0}^{L-{\cal N}}
\sum_{\{m_a\}} \,
d_E\, (\lambda_C)^C (\lambda_1)^{N_1} (\lambda_2)^{N_2} e^{-E/T}$$
where ${\cal N}=N_1+N_2$ is the total number of exclusons,  
$E$ is given in (\ref{Eexc}) and the sum 
$\sum_{\{m_a\}}$ is over sets of integers obeying $1\le m_1<m_2<\dots
<m_{\cal N}\le L$. Note that the maximum number of exclusons is
$L$, so if $f_1+f_2>L$, then $C\ge f_1+f_2-L$. 

One can find a Hamiltonian with the same grand canonical partition
function by considering spin-$1/2$ fermions with {long-range}
interactions. Having two fermions of different spin and the same
momentum corresponds to a Cooper pair. To make the Cooper pair have
energy zero, the Hamiltonian must include a binding energy for
fermions of the same momentum. Such a Hamiltonian is simple to write
in momentum space, but when Fourier-transforming back to position
space, one obtains a complicated long-range interaction.

Before proceeding to the solution by Bethe ansatz we comment once
again on the fractional statistics aspects of the model. We already
observed that particles on a given chain act as fermions, while
moving a fermion on one chain past one on the other involves phase
factors of $\pm i$. This then suggests a statistics matrix
(in the sense of Haldane \cite{Haldane91b}) $G=\{g_{ij}\}$ with 
$g_{11}=g_{22}=1$, $g_{12}=g_{21}=1/2$ for
these `bare' particles. For the exclusons, which fully exclude
one another from a given level, the natural assignment would be 
$g_{11}=g_{22}=1$, $g_{12}=g_{21}=1$. We observe though that the 
1-level partition sum that is associated with this choice of $G$
(using results of \cite{FuKa,BS}) is different from the simple 
result that we find here.

\section{The solution using the nested Bethe ansatz}
\label{sec:bethe}

In this section we derive all the claims made in the last section by
using the Bethe ansatz.

When all fermions are on the same chain, there are no interactions
save forbidding double-occupancy. Exact eigenstates are given in
(\ref{phifree}) above. To forbid double occupancy and anti-symmetrize
them appropriately, we introduce a permutation $P=(P1,P2,\dots,
Pf)$ of the integers $(1,2,\dots,f)$, where $f$ is the total
number of fermions. Then we have
\begin{equation}
\varphi_{(f,0)}=\sum_P (-1)^{|P|} \prod_{a=1}^{f} \sin(p^{}_{Pa} j_a)\ .
\label{phif0}
\end{equation}
where $|P|$ is the order of the permutation, and $p_a$ satisfies
(\ref{pfree}).

The Bethe ansatz is a very natural generalization of (\ref{phif0}). 
In situations like ours where there is more than one species of
particle, imposing Fermi statistics does not automatically fix the
relative coefficients between different orderings. One must therefore
use the {\em nested Bethe ansatz} \cite{Yang67}. It is thus
necessary to label the orderings by another permutation $Q$ of
$(1,2,\dots f)$, so that $\varphi^Q$ is the part of the
wavefunction where $j_{Q1}<j_{Q2}<\dots<j_{Qf}$.
It is simplest to first
work with periodic instead of open boundary conditions. This amounts
to changing the sums in $H_f$ in (\ref{Ham}) and $H_c$ in (\ref{Hc})
to run from $1$ to $L$, interpreting positions mod $2L$, and setting
$H_b=0$. The Bethe ansatz for the eigenstate with periodic boundary
conditions is then
\begin{equation}
\varphi^Q=\sum_P A_P^Q \exp\left(i\sum_{a=1}^f p^{}_{Pa}j_{Qa}\right).
\label{ansatz}
\end{equation}
For open boundary conditions, the eigenstates consist of sums over
these $\varphi^Q$ with $\pm p_a$, i.e.\ are standing waves. Note that
the momenta are permuted over all $f=f_1+f_2$ fermions.

The miracle of the Bethe ansatz in general is that in some situations,
the $p_a$ and the $A_P^Q$ can be found which result in an eigenstate of
the Hamiltonian. The corresponding energy is exactly that given in
(\ref{Ep}). The $p_a$ are given as solutions of what are called
the {\em Bethe equations}.  The additional miracle here is that for
open boundary conditions with (\ref{bcmag}), the Bethe equations can
be solved explicitly, yielding the results described in section
\ref{sec:results}.

Let us first discuss the case of two fermions. If both are on the same
chain, the only constraint is no double occupancy,
i.e. $\varphi^Q(j,j)=0$ for all $Q$. This means that
\begin{equation}
A^Q_{12}=-A^Q_{21} \qquad\quad \hbox{for }f_1=2,\ f_2=0,
\label{constraint0}
\end{equation}
for both $Q=(12),(21)$. The solution is much more interesting when
each chain has one fermion. To make the equations look simpler, let
$x=e^{ip_1}$ and $y=e^{ip_2}$, and $j$ and $k$ represent the locations
of the fermions on chains 1 and 2 respectively; by our conventions
$j$ is an odd integer and $k$ even.  so we have
\begin{eqnarray*}
\varphi^{12}(j|k) =& A_{12}^{12} x^{j} y^{k} + A^{12}_{21} y^{j} x^{k}\qquad
&\hbox{for } j<k\\
\varphi^{21}(j|k) =& A_{12}^{21} x^{k} y^{j} + A_{21}^{21} y^{k} x^{j}\qquad
&\hbox{for }k<j\ .
\end{eqnarray*}
We then let the Hamiltonian act on the state
(\ref{estates}) with this $\varphi$.  Ignoring the boundary conditions
momentarily, a little algebra shows that this is an eigenstate if
\begin{equation}
2\varphi^{Q}(k-1|k)=-\varphi^{Q}(k+1|k)-i\varphi^{Q'}(k+1|k)
-\varphi^{Q}(k-1|k-2)+i\varphi^{Q'}(k-1|k-2)
\label{constraint1}
\end{equation}
and its parity conjugate
\begin{equation}
2\varphi^{Q'}(k+1|k)=-\varphi^{Q'}(k+1|k+2)-i\varphi^{Q}(k+1|k+2)
-\varphi^{Q'}(k-1|k)+i\varphi^{Q}(k-1|k)
\label{constraint2}
\end{equation}
for all even $k$. Here $Q=12$ and
$Q'=21$, but it is straightforward to show that these constraints
apply in general when $Q'$ is defined by reversing $Qa$ and $Q(a+1)$
in $Q$, i.e.\ $Q'=(\dots Q(a-1),Q(a+1),Qa,Q(a+2),\dots)$. Then
(\ref{constraint1}) and (\ref{constraint2}) must hold whenever
$Qa\le f_1$ and $Q(a+1)>f_1$, i.e.\ $j_{Qa}$ is odd (is on chain 1), and 
$j_{Q(a+1)}$ is even (is on chain 2). 

The eigenvalue for
such an eigenvector is simply (\ref{Ep}), i.e.\
\begin{equation}
E_{(1,1)}=4+x^2+x^{-2}+y^2+y^{-2}=(x+x^{-1})^2+(y+y^{-1})^2. 
\label{E11}
\end{equation}
Plugging the Bethe ansatz (\ref{ansatz}) for the eigenvector into the
constraints (\ref{constraint1},\ref{constraint2}) 
gives two equations for the four unknowns $A_P^Q$. We find
\begin{equation}
{E_{(1,1)}}\begin{pmatrix} A^{12}_{21} \\ A^{21}_{21} \end{pmatrix}
=\begin{pmatrix} -2 (x+x^{-1})(y+y^{-1}) & i ((x+x^{-1})^2-(y+y^{-1})^2)\\
i( (x+x^{-1})^2-(y+y^{-1})^2) & -2 (x+x^{-1})(y+y^{-1})\end{pmatrix}
\begin{pmatrix} A^{12}_{12} \\ A^{21}_{12} \end{pmatrix}
\label{cbbc}
\end{equation}
When $(y+y^{-1})=\pm i(x+x^{-1})$, we have $E_{(1,1)}=0$, and the
particles form a Cooper pair. Since the matrix has determinant 
$(E_{(1,1)})^2$, it is still possible to solve (\ref{cbbc}) for
a Cooper pair, yielding $A^{12}_{12}=\pm A^{21}_{12}$ and
$A^{12}_{21}=A^{21}_{21}=0$. 

This eigenstate must also satisfy the boundary conditions. For
periodic untwisted boundary conditions, they are simply 
$$(xy)^L=1, \qquad y^L A_{12}^{12}=A_{21}^{21}, \qquad
x^L A_{12}^{21}=A_{12}^{21}\ .$$
The first of these is simply the requirement that the total momentum be
an integer multiple of $2\pi/L$, while the others are the individual
momentum quantization conditions in the presence of 
interactions. These equations can be solved to yield all of the
eigenstates. The Cooper pairs do not occur here, since to satisfy both 
the periodic boundary
conditions and (\ref{cbbc}) would require all $A_P^Q=0$. 

The generalization of this computation to arbitrary numbers of
particles is now standard, even though it is a mere 39 years since the
invention of the nested Bethe ansatz. The key ingredient is the
$R$-matrix, which encodes the relations on the $A_P^Q$ necessary to
make the state an eigenvector. These relations are
given in (\ref{constraint0},\ref{constraint1},\ref{constraint2}): one
must satisfy the constraints for each pair of fermions when they
become adjacent. Let $P=(\dots Pa,P(a+1) \dots)$
and $P'=(\dots P(a+1),Pa \dots)$. Then
we have
\begin{equation}
E_{Pa,P(a+1)} A_{P',Q} = R(p^{}_{Pa},p_{P(a+1)^{}})
A_{P,Q}
\label{ARA}
\end{equation}
where
\begin{equation}
R(p_a,p^{}_b)= 
\begin{pmatrix} 
-E_{ab}&0&0&0\\
0&-8\cos(p_a)\cos(p_b)&
{4i}(\cos^2(p_a) - \cos^2(p_b))&0\\ 
0&4i(\cos^2(p_a) - \cos^2(p_b))&
-{8}\cos(p_a)\cos(p_b)&0\\ 
0&0&0&-E_{ab}\\
\end{pmatrix}
\label{Rmat}
\end{equation}
where $E_{ab}$ is the energy of particles with momenta $p_a$ and
$p_b$, i.e.\
$$E_{ab}=4(\cos^2(p_a) + \cos^2(p_b)).$$ The $R$-matrix is $4\times 4$
because there are four possibilities for which chains the fermions at
locations $j_{Qa}$ and $j_{Q(a+1)}$ occupy. The $2\times 2$ block in
the middle applies to the cases where the fermions are on different
chains, and is the same as that in (\ref{cbbc}). 
The inverse $(R(p_a,p_b))^{-1}=R(p_b,p_a)/(E_{ab})^2$, as
necessary for consistency.  We have followed \cite{Baxter} and
written (\ref{ARA}) so that it still is applicable when
$E_{ab}\to 0$, i.e.\ when $p_a$ and $p_b$
satisfy the Cooper pair condition (\ref{pcp}).

The $R$ matrix relates coefficients $A^Q_{P'}$ to
$A^Q_{P}$. Repeatedly applying (\ref{ARA}) gives all the $A_P^Q$ in terms of
those for a given permutation, which we label as $P=0$. For
periodic boundary conditions the $A_0^Q$ are then related to each other by
imposing the boundary conditions as well.  To derive the resulting
Bethe equations in situations like ours where the $R$ matrix is
non-diagonal, one must do a second Bethe ansatz, the ``nested'' part
of the nested Bethe ansatz \cite{Yang67}. Here this requires one level
of nesting, and we omit it here because we will not need it. A
vital consistency condition in applying (\ref{ARA}) is that the
$R$-matrix satisfy the Yang-Baxter equation, which ensures that this
procedure defines all the $A_{P}^Q$ consistently.  This $R$-matrix
(\ref{Rmat}) indeed satisfies the Yang-Baxter equation, and for
example arises in vertex models \cite{Perk}
and the $t-J$ model with the $SU(2)$ symmetry deformed to
$SU(2)_q$ \cite{Karowski}. It also arises in the scattering matrix of the
kinks in the sine-Gordon model at its supersymmetric point
$\beta^2=16\pi/3$ \cite{ZZ}. We note also that this $R$-matrix
satisfies a special condition called the ``free-fermion'' condition
\cite{Baxbook}, which is part of the reason behind the miracles in
this paper.

We now return to the boundary conditions of interest in this paper:
open ones with a boundary magnetic field (\ref{bcmag}). 
To implement these open boundary conditions, we need to take combinations of
wavefunctions with $\pm p_a$. 
The Bethe
ansatz (\ref{ansatz}) is generalized here to
\begin{equation}
\varphi^Q=\sum_P A_P^Q \prod_{a=1}^f
\left(e^{i p_{Pa}j_{Qa}}- \gamma_{Pa}^Q e^{-i p_{Pa}j_{Qa}}
\right).
\label{ansatzopen}
\end{equation}

The boundary conditions can be satisfied by choosing the coefficients
$\gamma_{Pa}^Q$ for the momentum corresponding to the leftmost and
rightmost fermions on each chain. 
By definition of $Q$, the leftmost fermion is located
at $j_{Q1}$, while the rightmost one is located at $j_{Qf}$. 
For each $Q$, the form of the boundary conditions depends on
which chain the leftmost and rightmost fermions are on. We have
\begin{eqnarray*}
0=&\begin{cases}
\varphi^Q(j_{Q1}=1,\dots|\dots) + \varphi^Q(j_{Q1}=-1,\dots|
\dots)&\quad\quad\hbox{if }Q1\le f_1\\
\varphi^Q(\dots|j_{Q1}=0,\dots) &\quad\quad\hbox{if }Q1>f_1
\end{cases}&
\\
0=&
\begin{cases}
\varphi^Q(\dots|\dots,j_{Qf}=2L) 
+ \varphi^Q(\dots|\dots,j_{Qf}=2L+2)& \hbox{if }Qf>f_1\\
\varphi^Q(\dots,j_{Qf}=2L+1|\dots) & \hbox{if }Qf\le f_1
\end{cases}&
\end{eqnarray*}
Plugging (\ref{ansatzopen}) into these boundary conditions, we find
the same constraint for either condition. We have for the left end
\begin{equation}
\gamma_{P1}^Q=-1
\label{bc1}
\end{equation} 
for all $Q$. Likewise, whichever boundary condition for the right end is
applicable, we find
\begin{equation}
e^{-ip^{}_{Pf}2(2L+1)}\gamma_{Pf}^Q=-1.
\label{bc2}
\end{equation}

First let us consider the case where $E_{ab}\ne 0$ for any choice of
$a,b$, i.e.\ there are no Cooper pairs. Deriving the Bethe equations
and the relations among the coefficients $A_P^Q$ now proceeds in the
same fashion as for periodic boundary conditions. In fact, the
computation is virtually identical because the $R$-matrix is
independent of the signs of $p_a$ and $p_b$. Thus we can expand out
the product in the ansatz (\ref{ansatzopen}), and apply the same
relation (\ref{ARA}) to any of the resulting $2^{f}$ terms for a given
$A_Q^P$. Consistency then requires that all the $\gamma_{Pa}^Q$ for a
given $Q$ to be the same. Since one of them is fixed to be $-1$ by the
boundary conditions at the left end, this means
$\gamma_{Pa}^Q=-1$ for all $P$, $a$, and $Q$. We still need to satisfy
the boundary conditions (\ref{bc2}) at the right end. Since these must
hold for all permutations $P$ and $Q$, the only way this is possible
is if
\begin{equation}
e^{ip_{a}^{}2(2L+1)}=1
\label{momquant}
\end{equation}for all $a$. 
This is precisely the momentum
quantization condition for a free fermion
$$p_a= m_a \frac{\pi}{2L+1}$$
given in (\ref{pfree})!

Another miracle is that there are multiple states with the same energy:
fixing the $p_a$ does {\em not} fix all the $A_P^Q$.
Because the boundary conditions are the same for all $Q$,
they do not restrict the $A_P^Q$ at all.  The relation
(\ref{ARA}) can be used to relate the $A_P^Q$ to those for a given
permutation. Labeling this given permutation by $P=0$,
(\ref{ARA}) determines the other $A_P^Q$ in terms of $A_0^Q$, but it
does not fix the latter. 
The only relation between the different $A_0^Q$ is therefore that
required by Fermi statistics for the fermions on each chain. Fermi
statistics for chain $1$ means that when both $Qa\le f_1$ and $Qb\le
f_1$, we have $A_P^Q=-A_P^{Q'}$, where $Q'$ differs from $Q$ by
swapping $Qa$ and $Qb$, i.e.\ if $Q=(\dots,Qa,\dots,Qb,\dots)$ then
$Q'=(\dots,Qb,\dots,Qa,\dots)$. Fermi statistics for chain 2 means
that when both $Qa>f_1$ and $Qb>f_2$, $A_P^Q=-A_P^{Q'}$ as well. 
For a given set of momenta satisfying
(\ref{momquant}), there are therefore $\binom{f_1+f_2}{f_1}$ undetermined
coefficients in the Bethe ansatz. All of these are perfectly valid
eigenstates, and they all have the same energy. We thus recover the
degeneracy $d_E$ in (\ref{degen}) in the special case where there are
no Cooper pairs. 

The natural way of interpreting these results is for the
quasiparticles to be the exclusons described in section
\ref{sec:results}.  These quasiparticles have a statistical
interaction, because each momentum $p_a$ must be different: the
wavefunction vanishes if $p_a=p_b$ for $a\ne b$, because
$R(p_a,p_a)=-1$.  Moreover, even though fermions cannot hop between
chains, the wavefunction is not a product of separate factors for each
chain. A given momenta is not associated with one chain or the other:
the sum over $P$ includes permutations over all $f=f_1+f_2$ momenta.
These facts are exactly what happens with two species of
quasiparticles obeying exclusion statistics.

We now can see explicitly that we must include Cooper pairs to make
the Bethe ansatz complete. There are $\binom{L}{f}$ choices of momenta,
each with degeneracy $\binom{f}{f_1}$. However, 
$$\binom{L}{f_1}\binom{L}{f_2}> \binom{L}{f}\binom{f}{f_1},$$ so there
are more states than those involving solely exclusons. The derivation
that all momenta must obey the free-particle quantization condition
(\ref{momquant}) is valid when $E_{ab}\ne 0$ for all $p_a$ and
$p_b$. Therefore the missing states must include at least one Cooper
pair, a state where $E_{ab}=0$ for at least one pair $(a,b)$.

So let us first return to the case where $f_1=f_2=1$, and now let
$p_1$ and $p_2$ satisfy the Cooper-pair condition (\ref{pcp}), so that
$E=0$. As noted above, the requirements
(\ref{constraint1},\ref{constraint2}) are satisfied here if
$A_{12}^{12}=A_{12}^{21}$, and $A_{21}^{12}=A_{21}^{21}=0$. 
The wavefunction of
a single Cooper pair is therefore
\begin{equation}
\varphi_{C}(j|k) = 
\begin{cases}
A(x^{j}-x^{-j})(y^{k}-y^{4L+2-k})
&j<k\\
A(x^{k}-x^{-k})(y^{j}-y^{4L+2-j}) &k<j
\end{cases}
\label{phiC}
\end{equation} 
where $(y+y^{-1})=i(x+x^{-1})$. One can check directly that this
yields an eigenstate of $H_f+H_c$ with eigenvalue $0$ for any value of
$x$. Note that the momentum quantization condition (\ref{momquant}) does
{\em not} apply to the momenta in Cooper pairs.

To show that all the missing states are given by Cooper pairs, we
count how many linearly independent eigenstates (\ref{phiC})
yields. It is useful to define the new variable $z=(x+x^{-1})=i(y+y^{-1})$. 
Choosing 
$$A=\frac{y^{-(2L+1)}}{(x-x^{-1})(y-y^{-1})}$$ gives 
$\varphi_C$ as a polynomial in $z^2$ (it can be expressed in terms
of Chebyshev polynomials explicitly if desired). It is
simple to then see that the highest order of $z$ appearing in these
polynomials is $z^{2(L-1)}$, and the lowest is $z^0$. Since $z^2$ is
arbitrary, it follows that the dimension of the space of states of a
single Cooper pair is $L$. This agrees with the general degeneracy
formula for $C=1$ and no exclusons, $N_1=N_2=0$. This gives all the
missing states for $f_1=f_2=1$: there are $L(L-1)$ two-excluson
states, giving the correct total of $L^2$. Thus we have the nice result
that Cooper pairs make up the states missing due to imposing the
exclusion statistics.

Now we consider the general case. It is important to note that we do
not single out two fermions and call them a Cooper pair. Rather a
Cooper pair occurs when two {\em momenta} obey the condition
(\ref{pcp}). The relation (\ref{ARA}) still is sufficient to satisfy
the eigenvector requirements (\ref{constraint1},\ref{constraint2});
(\ref{constraint0}) is satisfied as long as we divide out any
$E_{ab}$ from both sides before taking them to zero.

When Cooper pairs are present, (\ref{ARA}) means that
some of the $A_P^Q$ vanish, and others must be equal. For $C$ Cooper pairs, 
we order the $p_a$ so that $\cos(p_{2c-1})=i\cos(p_{2c})$, for
$c=1,\dots C$. Given a permutation $P$, define $a_1$ and $a_2$
so that $Pa_1=2c-1$ and $Pa_2=2c$. 
Then for $A_P^Q$ not to vanish, we find that for all $c$
\begin{enumerate}
\item $(2c-1)$  must appear before $(2c)$ in $P$, i.e.\ $a_1<a_2$ 
\item if there are no fermions between those with the Cooper pair
  momenta (i.e.\ $a_2=a_1+1$) then these two particles must be on
  different chains. This means $Qa_1\le f_1$  and $Qa_2=Q(a_1+1)>f_1$,  
  or vice versa.
\end{enumerate}   
In the latter case, the $R$-matrix relation (\ref{ARA}) also requires that 
$A_P^Q=A_P^{Q'}$, where $Q'$ is $Q$ with $a_1$ and $a_2=a_1+1$ reversed. 

Applying (\ref{ARA}) means that the free-particle momentum
quantization condition (\ref{momquant}) is still applicable for all
the momenta not part of Cooper pairs, i.e.\ $p_a$ for $a>2C$. However
the momenta in Cooper pairs need not satisfy it. The reason is as
follows.  Because $Pa_1=2c-1$ must always appear before $Pa_2=2c$ in a
$P$ with a non-vanishing $A_P^Q$, the fermion at position $j_{Qa_1}$
must always be to the left of a fermion at position $j_{Qa_2}$. Thus
no fermion with momentum $p_{2c}$ can ever be the leftmost particle,
and no fermion with momentum $p_{2c-1}$ can ever be
rightmost. Therefore, the boundary condition (\ref{bc1}) does not
apply to any $\gamma^Q_{2c}$, and the boundary condition (\ref{bc2})
does not apply to any $\gamma_{2c-1}^Q$. Since (\ref{momquant}) arose
from demanding that the two boundary conditions on a given momentum be
consistent, it does not apply to any Cooper-pair momentum.

Let us illustrate these conditions with $f_1=2$ and $f_2=1$, where
there are $6$ different permutations for $P$ and three for $Q$ to
consider. (We only need consider $(123)$, $(132)$, and $(312)$ for $Q$
because the other three are given by applying Fermi statistics to the
two fermions on chain 1.) With no Cooper pairs, there are three
independent $A_{123}^Q$, as discussed before. For one Cooper pair, we
have $p_1$ and $p_2$ satisfying (\ref{pcp}). This means that when
$\widetilde{P}=(213),(231),(321)$, $A_{\widetilde{P}}^Q$ vanishes for
all $Q$. When $P=(123)$, we have $a_1=1$ and $a_2=2$, so this means
$A_{123}^{123}=0$ as well. The non-zero ones must obey for example
$A_{123}^{132}=A_{123}^{312}$. Thus for the three $A_{123}^Q$, one is
zero and the other two are equal. Applying (\ref{ARA}) then determines
all the other $A_P^Q$, and including the boundary conditions requires
that $p_3$ obey the free-fermion condition (\ref{pfree}). Therefore
all the $A_P^Q$ are given in terms of $A_{123}^{312}$: all the
degeneracies here arise from the arbitrariness of $p_1$. Counting the
different degrees of freedom here is similar to the two-particle
case. We replace $p_1$ and $p_2$ with $z$, so that the wavefunction is
given in terms of polynomials in $z$. However, because of the presence
of the excluson with momentum $p_3$, the lowest-order term in
$\varphi^{312}$ is now order $z^2$, so there are only $L-1$ terms in
the polynomial. When there are no other exclusons, there are $L$
terms. The reason for this excluded volume effect is Fermi
statistics. The wavefunction must vanish when $j_1=j_2$, so neither
$p_2$ nor $p_1$ can be $p_3$, thus reducing the number of
possibilities for the momentum. This yields a total $L(L-1)$ states
with one Cooper pair (the factor of $L$ coming from the $L$ possible
values of $p_3$). Likewise, there are $3\binom{L}{3}$ states with no
Cooper pairs, the factor of 3 coming from the independent
$A_{(123)}^Q$.  This is a complete set of states:
$$\binom{L}{1}\binom{L}{2}=3\binom{L}{3}+L(L-1)$$

To obtain the general degeneracy given in (\ref{degen}), one proceeds
in this fashion. The first factor comes from the number of independent
$A_0^Q$, while the second comes from the order of the polynomials in
$z_c$ describing the momenta in the $c$th Cooper pair. Let us first
discuss the different $A_0^Q$, where $0$ is our label for $(12\dots
f)$. Applying Fermi statistics to chain $1$ relates all the $Q$
obtained by exchanging any of $Qa$ with $Qa\le f_1$, while applying it
to chain $2$ relates any obtained by exchanging any of the $Qa$ with
$Qa>f_1$. Thus the largest possible number of distinct $A_0^Q$ is
$\binom{f}{f_1}$. We showed above that these are all distinct when
$C=0$. Now consider $C=1$.  The second condition for non-vanishing
$A_P^Q$ described above
means that $A_0^Q$ vanishes for any $Q$ where $j_{Q1}$ and $j_{Q2}$
are either both even or both odd. The non-vanishing ones therefore
have $Q1\le f_1$ and $Q2> f_1$ or vice-versa. There are
$2f_1f_2(f-2)!$ such possibilities. If $Q'$ is $Q$ with $Q1$ and $Q2$
reversed, then $A_0^{Q}=A_0^{Q'}$, so the number of distinct
possibilities is lowered to $f_1f_2(f-2)!$. Including the effects of
Fermi statistics as well means dividing by $f_1!$ and $f_2!$. Thus the
total number of distinct $A_0^Q$ for one Cooper pair is
$\binom{f-2}{f_1-1}$. For general $C$, applying the second condition
$C$ times in the analogous fashion gives
$$\frac{f_1!}{(f_1-C)!}\frac{f_2!}{(f_2-C)!}\frac{(f-2C)!}{f_1!f_2!}=
\binom{f-2C}{f_1-C}$$ 
for the number of independent $A_0^Q$. Thus we indeed recover the
first factor in (\ref{degen}). 

The second factor likewise follows from essentially the same argument
we gave for small $f_1$ and $f_2$. The presence of fermions with
momenta not obeying the Cooper-pair conditions creates an excluded
volume effect reducing the number of independent terms in the
polynomials in $z_c$. The Fermi statistics means that the wavefunction
must be symmetric under exchanges of the $z_c$, implying that the
number of linearly-independent choices of the $z_c$ is the binomial
$$\binom{L-(f-2C)}{C}.$$ By writing the grand canonical partition
function as at the end of section \ref{sec:results}, we saw that including
the degeneracy (\ref{degen}) successfully counts all the states in the
theory. Thus the Bethe ansatz is complete.

\section{The extended (super)symmetry algebra}
\label{sec:susy}

The extensive degeneracies in the spectrum given by (\ref{degen})
imply that our model must possess an extended symmetry algebra. This
is obvious in the quasiparticle basis of states in terms of exclusons
and Cooper pairs. A symmetry operator creates or annihilates a Cooper
pair, or changes the type of a excluson, preserving the momentum.  In
this section we describe this symmetry algebra, and how it acts on the
orginial fermionic states. The symmetry turns out to be an extension
of supersymmetry, and is infinite-dimensional as $L\to\infty$. As we
will show, it is quite reminiscent of the Yangian symmetry of the
Haldane-Shastry model.


The basic supersymmetry operators act on the exclusons:
$Q$ changes type 1 to type 2, and $Q^\dagger$
the inverse. Changing the
type of an excluson requires flipping a fermion from one chain
to the other. Thus $Q$ hops a particle on chain 1 to chain 2, with
$Q^\dagger$ doing the reverse. These operators are non-local: they
depend on the locations of the other particles. Let
$$\alpha_k \equiv \sum_{j=1}^{k} (-1)^j n_j.$$
We then define
\begin{equation}
Q=c_2^\dagger c_1 + \sum_{k=1}^{L-1} 
\left(e^{i\alpha_{2k-1}\pi/2} c^\dagger_{2k} + 
e^{i\alpha_{2k}\pi/2}c^\dagger_{2k+2}\right) c_{2k+1}.
\label{Qdef}
\end{equation}
Both $Q$ and $Q^\dagger$ annihilate the Cooper pairs.

This supersymmetry charge commutes with $H=H_f+H_c$ if $H_b$ is the
boundary magnetic field in (\ref{bcmag}). This can be verified
directly, but it is more illuminating to work out the entire
supersymmetry algebra. Even though $Q$ is written in terms of fermion
bilinears, it is effectively fermionic because of the non-local string
of operators from the exponential. One finds for its anticommutators
\begin{equation}
Q^2 = 0\ , \quad (Q^\dagger)^2=0\ , \quad H=\{Q,Q^\dagger\} \ .
\label{HQQ}
\end{equation}
It then follows from (\ref{HQQ}) that $[Q,H]=[Q^\dagger,H]=0.$

One can derive a number of interesting properties of the states
directly from the (unextended) supersymmetry algebra
(\ref{HQQ}). Since we have already derived them (and many more) by
using the Bethe ansatz, we will just state these properties here; see
\cite{FS} for detailed explanations of the methods. All eigenvalues
$E$ of the Hamiltonian obey $E\ge 0$. States with $E>0$ form doublets
under the supersymmetry algebra, while the $E=0$ ground states are
annihilated by both $Q$ and $Q^\dagger$.  The Witten index provides a
lower bound on the number of $E=0$ ground states
\cite{Witten82}. Since the supersymmetry generators are fermionic, for
purposes of computing the Witten index, each fermion on chain 1 has
charge $-1/2$, while each on chain 2 has charge $+1/2$. Then
$$W=\sum_{\hbox{states}} e^{i(f_2-f_1)\pi/2}=2^L,$$ so there are at
least $2^{L}$ ground states for all possible values of $f_1$ and $f_2$
for a given size $L$.  We showed above that a state with $E=0$ must be
comprised solely of Cooper pairs. These occur whenever $f_1=f_2$, and
from (\ref{degen}) we see that the number of such states for a fixed
$f_1$ and $f_2$ is $\binom{L}{f_1}$ because $C=f_1=f_2$ here. Thus
there are exactly $2^L$ ground states, and the lower bound from the
Witten index is the exact number here. One can find a Hamiltonian with
twisted periodic boundary conditions obeying (\ref{HQQ}). The Witten
index is the same, so the Cooper pairs remain for the twisted model,
but the other degeneracies no longer occur.

To understand the full degeneracies, supersymmetry alone is not
enough.  $Q$ does not
change the total fermion number, whereas annihilating or creating a
Cooper pair will annihilate or create fermions on both chains.
Starting from a state with all particles on the upper chain,
acting with $Q$ flips one particle to the lower chain.  Since $Q$ is
nilpotent the action cannot be repeated.  Moreover,  To
obtain the full degeneracies $\binom{f}{f_1}$ of the exclusons one
would need $f$ generalized supercharges that can act independently. We
have found that a hierarchy of operators $Q^+_p$, $Q^-_p$, $H_p$, with
$p=0,1,\ldots$, can be constructed with the following algebraic
properties
\begin{equation}
\{Q^+_p,Q^+_q\}=0, \quad
\{Q^-_p,Q^-_q\}=0, \quad
\{Q^+_p,Q^-_q\}=H_{p+q}\ ,
\label{syangian}
\end{equation}
implying that $[Q_p^\pm, H_q]=0$ for all $p,q$. We refer
to this algebra as a `super-Yangian', by analogy with the Yangian
appearing in the Haldane-Shastry chain \cite{HSYangian}.

The leading terms in the hierarchy are $Q_0^+=Q$, $Q_0^-=Q^\dagger$, 
$H_0=H_f+H_b+H_c$. For the operator $Q_1^+$ we choose a form where a 
particle at site $i$ in the upper chain hops to sites $i+j$, $j=1,3,\ldots$
in the lower chain with amplitudes containing an alternating sign 
$(-1)^{(j-1)/2}$,
\begin{equation}
Q_1^+= \sum_{k=0}^{L-1} e^{-i\alpha_{2k}\pi/2}
 \left( \sum_{j=1,3,\ldots,2N-2k-1} 
 (-1)^{(j-1)/2} e^{i\beta_{2k,j} \pi/2} c^\dagger_{2k+1+j} \right) c_{2k+1} \ ,
\label{Q1def}
\end{equation}
with $\alpha_{2k}$ as given above and 
$$\beta_{2k,j} \equiv 2 (n_{2k+2}+n_{2k+4}+\ldots+n_{2k+j-1})-
                       (n_{2k+3}+n_{2k+5}+\ldots+n_{2k+j}) \ .$$
On a 1-particle Bethe state with $x=e^{ip}$ these amplitudes combine as
$$ x^{-1} - x^{-3}+ x^{-5} - \ldots \rightarrow (x+x^{-1})^{-1}. $$
As a consequence, we find that on a general 1-particle Bethe state the 
operators $H_p$ act diagonally with eigenvalue
\begin{equation}
E_p[x] = (x+x^{-1})^{2-2p} \ .
\end{equation}
On a general $f$-particle state we find that $H_1=0$ for $f$ even 
while $H_1=1$ on all states with $f$ odd.


We have investigated the explicit form of $Q_p^\pm$ with $p\geq 2$
and $H_q$ with $q\geq 3$ for small system sizes, confirming that 
operators satisfying the full algebra eq.~(\ref{syangian}) indeed
exist. Restricting to 2-particle states, there is some freedom to 
choose the $Q_p^\pm$; what seems to be a canonical choice leads 
to eigenvalues 
\begin{equation}
E_p[x,y] = (x+x^{-1})^{2-2p} + (-1)^p  (y+y^{-1})^{2-2p}
\end{equation}
on states with two exclusons. In addition, the $H_p$ have an 
$L$-fold degenerate eigenvalue $E_p=0$ when acting on states
with a single Cooper 
pair. States with more particles show a similar pattern;
for $f$ particles with $C$ Cooper pairs the eigenvalues $E_p$
reduce to those of $f-2C$ exclusons. The operators can be chosen 
such that on a state with exclusons at $\{x_a\}$, $a=1,\ldots,f$,
the eigenvalues of $H_p$ with $p$ even take the form
\begin{equation}
E_p[\{x_a\}] = \sum_{a=1}^f (x_a+x_a^{-1})^{2-2p}
\qquad {\rm for}\ p \ {\rm even.} 
\end{equation}

The algebraic structure in eq.~(\ref{syangian}) is highly reminiscent
of the Yangian $Y(sl_2)$ which features as the symmetry algebra
of various integrable models \cite{Mackay}. In the Haldane-Shastry 
spin-chain, the $SU(2)$-spin symmetry, with generators $Q^A_0$ with
 $A={\pm,3}$, extends to a Yangian generated by $Q^A_p$, $p=0,1,\ldots$, 
with all $Q_p^A$ commuting with conserved quantities $H_q$ 
\cite{HSYangian}. This algebraic 
structure beautifully reflects the underlying physical picture of 
`spinons', with the Yangian generators $Q_p^\pm$ performing spin-flips 
on a multi-spinon state with all spins polarized. In our model here,
the situation is similar, with the $Q_p^\pm$ sweeping out a full 
multiplet of eigenstates from a reference state with all particles 
`polarized' on the same chain. Comparing with the Haldane-Shastry model,
the role of the $SU(2)$-spin symmetry is taken over by supersymmetry,
bringing the charges $Q_p^\pm$ and the conserved quantities $H_q$
together in one algebraic structure. The actual form of the
first nontrivial `sYangian' generators $Q_1^\pm$ is reminiscent of
the form of the (bosonic) Yangian generators (sometimes referred
to as the `logarithmic Yangian') in integrable quantum field theories
of massive particles and WZW models of conformal field theory 
\cite{logyangian}.

The results of this section apply to open boundary
conditions with the boundary magnetic field (\ref{bcmag}). An
interesting open question is if this symmetry algebra can be deformed
to apply to other boundary conditions, where the model is still
solvable but there no longer exist the large degeneracies
(\ref{degen}). If such a symmetry algebra does persist, it is then
spectrum-generating like the Yangian.

We remark that in addition to all symmetries already mentioned, 
the model (with open boundary conditions) admits a discrete symmetry 
reminiscent of that of the Hubbard model at half filling. It is implemented
by interchanging particles and holes {\it on the upper chain only},
leading to
\begin{equation}
(f_1,f_2) \leftrightarrow (L-f_1,f_2), \qquad E_0 \leftrightarrow 2L-1-E_0,
\end{equation}
and in particular
\begin{equation}
N_2 \leftrightarrow C \ .
\end{equation}
The latter relation shows that the particle-hole transformation 
maps the operators $Q_p^+$ (which each flip one particle from
upper to the lower chain) into operators creating Cooper pairs.

\section{Conclusion}
\label{sec:conclusion}

We have discussed in depth a one-dimensional model with local
interactions whose excitations are Cooper pairs and quasiparticles
obeying exclusion statistics. We used the Bethe ansatz to derive these
results, and found that various technical complications ended up being
crucial to describing some very interesting physics.

Despite the presence of Cooper pairs, our system is not really a
superconductor, because it is gapless. However, we believe it is
possible to modify the Hamiltonian to give the exclusons a gap while
preserving the Cooper pairs. The reason is the supersymmetry. If we
deform the couplings without breaking the supersymmetry, the Witten
index will not change \cite{Witten82}, so the $2^L$ ground states
with $E=0$ will remain. Any supersymmetry-preserving deformation which
gaps the exclusons will therefore leave the Cooper pairs. The next
step would be to break the supersymmetry slightly so that Cooper pairs
can have a kinetic energy below the gap.

Our model has some intriguing similarities to a solvable
one-dimensional model of a superconductor.  The ``Russian-doll'' model
is an effective model of Cooper pairs, where time-reveral symmetry is
broken by including a complex phase in the Hamiltonian
\cite{Russiandoll}.  This is very reminiscent of how time-reversal
symmetry is broken in our model, and it would be interesting to find a
deformation of our model whose Cooper pairs are described by the
Russian-doll model.

We believe our model will prove useful in understanding exclusion
statistics as well as superconductivity, since it can be treated with
well-studied Bethe-ansatz methods. It should be possible to derive
many more interesting properties for it. We are also hopeful that it
may shed light on anyonic superconductivity.  We find it amazing that
75 years after the invention of the Bethe ansatz, it is still being
used to derive and explore fascinating properties of one-dimensional
quantum systems.

\bigskip\bigskip

We are very grateful to Fabian Essler for his patience.
This work was supported in part by the foundations FOM and
NWO of the Netherlands, and by NSF grant DMR-0412956.

\end{document}